\documentclass[amsmath,amssymb, reprint,numerical]{revtex4-1}

\usepackage{graphicx}% Include figure files
\usepackage{dcolumn}% Align table columns on decimal point
\usepackage{bm}% bold math
\usepackage{graphics, caption, subcaption}% Include figure files
\usepackage[utf8]{inputenc}
\usepackage[T1]{fontenc}
\usepackage{amsmath,booktabs,braket,cleveref,mathptmx,xcolor,siunitx}
\usepackage[version=4]{mhchem} 
 \usepackage[normalem]{ulem}
\usepackage{array}
\usepackage{paralist}
\usepackage{multirow}

\usepackage{chemformula}
\usepackage{bm}

\newcommand{\abinitio}{\emph{ab initio}}
\newcommand{\cm}{cm$^{-1}$}

\newcommand{\totalsf}{1495}

\def\a0{{$a_{\rm 0}$}}

\newcommand{\mc}{\multicolumn}
\newcolumntype{H}{>{\setbox0=\hbox\bgroup}c<{\egroup}@{}}

\newcolumntype{d}{D{.}{.}{-1}}

\usepackage{float}

\usepackage{natbib} % Last name et al. references
\usepackage{threeparttable} % Footnotes in table

\begin{document}

Article Type: Focus Article

Corresponding Author: Dr Laura K. McKemmish, UNSW Sydney (l.mckemmish@unsw.edu.au)

\newpage

\title{Meta-analysis of Uniform Scaling Factors for Harmonic Frequency Calculations}% \ce{C2}: Updating collated datasets with new experiment} 
%Effect of basis set on dipole moment.
%Data collation in molecular diatomic spectroscopy: An update to the $C_2$ MARVEL compilation and ExoMol line list, and a discussion of molecular constants for nine diatomics \\}

\author{Juan C. Zapata Trujillo}
\affiliation{School of Chemistry, University of New South Wales, 2052 Sydney}

\author{Laura K. McKemmish}
\email{l.mckemmish@unsw.edu.au}

\affiliation{School of Chemistry, University of New South Wales, 2052 Sydney}

\date{\today}

\begin{abstract}
Vibrational frequency calculations performed under the harmonic approximation are widespread across chemistry. However, it is well-known that the calculated harmonic frequencies tend to systematically overestimate experimental fundamental frequencies; a limitation commonly overcome with multiplicative scaling factors.

In practice, multiplicative scaling factors are derived for each individual model chemistry choice (i.e., a level of theory and basis set pair), where performance is judged by, for example, the root-mean square error (RMSE) between the predicted scaled and experimental frequencies. However, despite the overwhelming number of scaling factors reported in the literature and model chemistry approximations available, there is little guidance for users on appropriate model chemistry choices for harmonic frequency calculations. 

Here, we compile and analyse the data for \totalsf{} scaling factors calculated using 141 levels of theory and 109 basis sets. Our meta-analysis of this data shows that scaling factors and RMSE approach convergence with only hybrid functionals and double-zeta basis sets, with anharmonicity error already dominating model chemistry errors. Noting inconsistent data and the lack of independent testing, we can nevertheless conclude that a minimum error of 25\,\cm{} -- arising from insufficiently accurate treatment of anharmonicity -- is persistent regardless of the model chemistry choice. Based on the data we compiled and cautioning the need for a future systematic benchmarking study, we recommend  $\omega$B97X-D/def2-TZVP for most applications and B2PLYP/def2-TZVPD for superior intensity predictions. With a smaller benchmark set, direct comparison strongly prefers $\omega$B97X-D/6-31G* to B3LYP/6-31G*.

\end{abstract}

\maketitle

\section{Introduction}
\label{sec:intro}

Predictions of fundamental vibrational frequencies by multiplicative scaling of computed harmonic frequencies is widespread across chemistry, as evidenced by the thousands of citations to papers computing recommended scaling factors (see for example \citet{96ScRa} and \citet{07MeMoRa}). Yet, central questions concerning harmonic frequency calculations have gone almost completely unaddressed; what model chemistry choices (i.e., a level of theory and basis set pair) should be used at different cost points, how important is this choice, and what errors can be expected in the predicted frequencies. 

Today, vibrational frequency calculations are one of the most routinely performed quantum chemistry calculations \cite{17PaFaRi,18Hs,19KuWa,19MaLeKa,19Oz,20TaOhSu,20PuBa,21ZaSyRo,21SpGr}. Arguably, their most common application is aiding in the interpretation of experimental vibrational spectra, where the computational predictions support accurate band assignments, especially for larger molecules with complex vibrational spectral patterns \cite{06KrZa,11PeJoTo,19GhUlMe,20Ke,21RuBr,21YuGaJi}. Nonetheless, the range of applications expands to the estimation of thermodynamic and kinetic properties such as zero-point energies \cite{88FlPoBa,02DyShYo,12FoHuFr,18SpGhLe,20CoHoQu}, entropic contributions to reaction energies and activation barriers \cite{08AgCuKl,17ZaGeDy,19KeBaSh,19HiBoLi,20DaHe}; as well as in the development of composite \abinitio{} methods such as the G$n$ \cite{89PoHeFo, 90CuJoTr, 91CuRaTr, 98CuRaRe, 07CuReRa} and W$n$ \cite{99MaOl,04BoOrAt, 06KaRaMa} approaches.

The harmonic approximation \cite{06Tu} is the preferred choice in most vibrational frequency calculations due to its relative computational simplicity, reliability and affordable scaling with larger systems \cite{97BaLa,14ThLaCh,17WaZhDo,18ScKuRy,20DaHe,21FoDe}. However, it is well-known that the calculated harmonic frequencies tend to systematically overestimate the experimentally-derived fundamental frequencies, limiting the accuracy in the predictions.

Different strategies involving scaling factors have been proposed to addressed the overestimation in the calculated harmonic frequencies (as reviewed by \citet{19Pa} and \citet{19BaBo}). However, global uniform multiplicative scaling factors (hereafter just uniform scaling factors) have become the most popular approach to produce scaled harmonic (predicted fundamental) frequencies that can be directly compared to experimental fundamental frequencies. Typically, this approach involves collating a reference set of experimental fundamental frequencies, $\nu_i$, and calculated harmonic, $\omega_i$, frequencies that are used to find an optimal scaling factor, $\lambda$, calculated as $\lambda = (\sum_{i}^{N} \omega_{i}\nu_{i})/(\sum_{i}^{N}\omega_{i}^{2})$, where both summations run over the total number of vibrational frequencies considered $N$. The set of calculated harmonic frequencies, $\omega_i$, is dependent upon the model chemistry choice, and therefore different values of $\lambda$ are calculated for each individual level of theory and basis set pair. Commonly, the quality of the frequencies after scaling is reported as root-mean-squared-error, i.e., $\textrm{RMSE} = \sqrt{\left( \sum_{i}^{N} (\lambda \omega_{i} - \nu_{i})^{2}) \right)/N}$.
%, where $N$ is the total number of vibrational frequencies considered.

We can attribute three main sources of error to the predicted fundamental frequencies (i.e., scaled harmonic frequencies): the lack of anharmonicity in the calculations, level of theory errors and basis set errors. 

The lack of anharmonicity in the calculations is irreducible within the harmonic approximation for a uniform scaling factor; using a benchmark data set of 119 frequencies from 30 molecules, \citet{15KeBrMa} estimate the anharmonicty error as 24.9\,\cm{} based on an optimal scaling of experimentally extracted harmonic frequencies against experimental fundamental frequencies. Noting that this value will be dependent on the benchmark data set, 25\,\cm{} nonetheless represents an approximate lower bound to the accuracy of predicted fundamental frequencies computed under the double-harmonic approximation. This error can be overcome by pursing full anharmonic calculations \cite{10ScLaBe, 13RoGe, 19PuBlTa}. However, while the double-harmonic approach requires only second-order derivatives of the potential energy, anharmonic treatments require higher-order derivatives and are thus far more computationally expensive, especially for larger molecules. 

The errors in the level of theory and basis set (aka model chemistry error) should be largely uncorrelated with the anharmonic error, and thus the goal is to determine a model chemistry that, at a affordable computational cost, has an accuracy close to 25\,\cm{}. Usually, an statistical figure (e.g., RMSE, MAD, uncertainty) is reported along with the scaling factor to quantify the quality expected in the scaled frequencies, and therefore this value can be used to approximately quantify the performance of different model chemistry choices in harmonic frequency calculations.

With the diversity of uniform scaling factors (and associated statistical figures) currently reported in the literature, the main goals of this paper are to: 

\begin{enumerate}
    \item compile all available uniform scaling factors data into a centralised repository (Section II);
    \item systematically review key developments in the construction of scaling factors, (Section III);
    \item perform a meta-analysis of the compiled scaling factors data, highlighting general trends between the scaling factors and their statistical figures, %, considering the trends and relationships between the scaling factors and their statistical figures 
    (Section IV);
    \item identify strong model chemistry choices for harmonic frequency calculations based on the existing data (Section V); and
    \item discuss promising future research directions (Section VI).
\end{enumerate}

We note that different but related scaling factors can also be used to reduce the error when predicting true harmonic frequencies and zero-point vibrational energies (ZPE). However, analysis on these types of scaling factors are beyond the scope of this paper. We encourage the reader to look into the Database of Frequency Scale Factors for Electronic Model Chemistries \footnote{https://comp.chem.umn.edu/freqscale/, date of access}, that provides tabulated scaling factors for the aforementioned properties together with references to the original publications.

\section{Consolidation and Overview of Uniform Scaling Factors Data} 
\label{sec:overview}

\begin{table*}[t!]
\centering
     \caption{Summary of literature review search of uniform scaling factors listing the number of scaling factors (\textit{SFs}), benchmark data set (\textit{Benchmark}), statistical figure (\textit{Stat.}), optimisation method (\textit{Method}) and number of basis set zeta-qualities and level of theory classes per publication. The column \textit{Dual} indicates whether the publication considers a combination of global and low- and high-frequency scaling factors (Y) or not (N). The basis set columns are organised in increasing zeta-quality. Likewise, the levels of theory classes are organised in increasing computational accuracy.}
     \label{tab:lit_review}
\scalebox{0.8}{
\begin{tabular}{lccclllcccccccccccccc}
    \toprule
    
    \multirow{2}{*}{Reference} & \multirow{2}{*}{Year} & \multirow{2}{*}{SFs} & \multirow{2}{*}{Dual} & \multirow{2}{*}{Benchmark$^{(a)}$} & \multirow{2}{*}{Stat.$^{(b)}$} & \multirow{2}{*}{Method$^{(c)}$} & \mc{5}{c}{Basis Set Zeta-Quality} & \mc{9}{c}{Level of Theory Class$^{(d)}$} \\
    \cmidrule(r){8-12} \cmidrule(r){13-21}
     &   & &   &   &   &   & \mc{1}{c}{SZ} & \mc{1}{c}{DZ} & \mc{1}{c}{TZ} & \mc{1}{c}{QZ} & \mc{1}{c}{5Z} & \mc{1}{c}{SE} & \mc{1}{c}{HF} & \mc{1}{c}{LSDA} & \mc{1}{c}{GGA} & \mc{1}{c}{mGGA} & \mc{1}{c}{H} & \mc{1}{c}{MP2} & \mc{1}{c}{DH} & \mc{1}{c}{P-MP2} \\ 
     
    \midrule
    \vspace{-0.8em} \\
    
    \textit{Global Data Sets$^{(e)}$} & & & & & & & & & & & & & & & & & & & & \\
             &     &   &              &      &    &   &    &    &    &   &   &   &   &    &   &    &   &   &    \\
    \citet{15KeBrMa} & 2015 &  308 & N  & 119f/30mol  & RMSE  & LS & - & 6 & 8  & 6 & - & - & - & - & 2 & 2 & 7 & 2 & 3 & 2  \\ 
    \citet{17KaChNe} & 2017 & 273 & N  & 99f/26mol   & RMSE  & LS & - & 2 & 15 & 4 & - & - & - & - & 1 & 3 & 9 & - & - & -  \\
    \citet{10AlZhZh} & 2010 &  199 & N    & 50f/15mol & RMSE$^{(f)}$  & LS                           & - & 15 & 24 & 2 & -& 5 & 1 & - & 13 & 10 & 39 & 2 & - & 15 \\
    \citet{07MeMoRa} & 2007 & 169 & Y & 1064f/122mol & RMSE  & LS & - & 3  & 2  & -  & - & - & 1 & - & 10 & 3 & 19 & 1 & - & 4  \\
    \citet{11LaBoHa} & 2011 &  90  & Y & 510f/42mol   & RMSE  & LS & - & 2  & 2  & 2  & - & - & - & - & 5  & - & 9  & - & 1 & -  \\
    \citet{12LaCaWi} & 2012 &  85  & Y & 510f/42mol   & RMSE  & LS & 1 & 1  & 1  & 1  & 1 & - & - & - & 6  & - & 10 & - & 1 & -  \\
    \citet{06TaPhRo} & 2006 & 72  & N & 1064f/122mol & RMSE  & LS & - & 6  & 3  & -  & - & - & - & - & 3  & 1 & 4 & - & - & -  \\
    \citet{21UnNaOz} & 2021 & 24  & N & 824f/157mol & Unc.  & LS & - & 15  & 9  & -  & - & - & - & - & - & - & 1 & - & - & -  \\
    \citet{14FrToHa} & 2014 & 24  & N & 1033f/117mol & RMSE  & LS & - & 1  & 2  & 1  & - & - & - & - & -  & - & - & 3 & - & 3  \\
    \citet{05IrRuRa} &2005 &  40  & N & 3939f/358mol & Unc. & LS & - & 4  & 1  & -  & - & - & 1 & - & 2  & - & 3  & 1 & - & 1  \\
    \citet{10TeMeCo} & 2010 & 30  & N & 1069f/143mol & Unc. & LS & - & 7  & 3  & -  & - & - & 1 & - & -  & - & 2  & - & - & -  \\
    \citet{17Cha}    & 2017 & 22  & Y & 1064f/122mol & None & LS & 4 & 4  & 1  & -  & - & 3 & - & - & 6  & - & 5  & - & - & -  \\
    \citet{96ScRa}   & 1996 & 19  & Y & 1064f/122mol & RMSE  & LS & - & 4  & 2  & -  & - & 2 & 1 & - & 2  & - & 3  & 2 & - & 1  \\
    \citet{04SiBoGu} & 2004 & 18  & Y & 510f/42mol   & RMSE  & LS & - & 2  & 2  & 2  & - & - & 1 & - & -  & - & 1  & 1 & - & -  \\
    \citet{08AnGoJo} & 2008 & 14  & Y & 312f/63mol   & RMSE  & LS & - & 1  & 1  & -  & - & - & 1 & - & 1  & - & 4  & 1 & - & -  \\
    \citet{16ChRa}   & 2016 & 9   & Y & 1064f/122mol & None & LS & - & -  & 5  & -  & - & - & - & - & -  & - & 1  & - & 5 & -  \\
    \citet{04WiMo}   & 2004 & 7   & N & 1303f/65mol  & MAD  & MM & - & 1  & -  & -  & - & 4 & 1 & - & 1  & - & 1  & - & - & -  \\
    \citet{05AnUv}   & 2005 & 7   & Y & 950f/125mol  & RMSE  & LS & - & -  & 7  & -  & - & - & - & - & -  & - & 1  & - & - & -  \\
    \citet{19Ha}     & 2019 & 7   & N & 1064f/122mol & RMSE  & LS & - & -  & 1  & -  & - & - & - & - & -  & - & 7  & - & - & -  \\
    \citet{01HaVeSc} & 2001 & 6   & Y & 900f/111mol  & Stand. Dev.  & LS & - & -  & 1  & -  & - & - & 1 & 1 & 1  & - & 2  & 1 & - & -  \\
    \citet{96Wo}     & 1996 & 6   & N & 1064f/122mol & RMSE  & LS & - & 1  & -  & -  & - & - & - & 1 & 2  & - & 2  & 1 & - & -  \\
    \citet{11DoBe}   & 2011 & 5   & N & 18f/18mol    & MAD  & LS & - & -  & -  & 1  & - & - & - & - & -  & - & -  & 5 & - & -  \\
    \citet{07FeHoKo} & 2007 & 2   & N & 922f/90mol   & RMSE  & LS & - & -  & -  & -  & - & 2 & - & - & -  & - & -  & - & - & -  \\
    \citet{93PoScWo} & 1993 & 2   & N & 1064f/122mol & RMSE  & LS & - & 1  & -  & -  & - & - & 1 & - & -  & - & -  & 1 & - & -  \\
    \citet{81PoScKr} & 1981 & 1   & N & 486f/83mol   & None & LS & - & 1  & -  & -  & - & - & 1 & - & -  & - & -  & - & - & -  \\
             &     &   &              &      &    &   &    &    &    &   &   &   &   &    &   &    &   &   &    \\
    \textit{Band-specific Data Sets$^{(g)}$} & & & & & & & & & & & & & & & & & & & & \\
             &     &   &              &      &    &   &    &    &    &   &   &   &   &    &   &    &   &   &    \\
    \citet{15AsDeBr} & 2015 & 12  & N & COf/20mol    & RMSE  & LS & - & -  & 3  & -  & - & - & - & - & 1  & 1 & 2  & - & - & -  \\
    \citet{08SiMaEh} & 2008 & 8   & N & COf/41mol & RMSE & LS & - & 1  & 1  & -  & - & - & - & - & 3  & - & 1  & - & - & -  \\
    \citet{03YuSrSc} & 2003 & 4   & N & C$\equiv$Of/31mol   & RMSE  & LS & - & 2  & -  & -  & - & - & 1 & - & -  & - & 1  & - & - & -  \\
             &     &   &              &      &    &   &    &    &    &   &   &   &   &    &   &    &   &   &    \\
    \textit{Total$^{(h)}$}    & &  \totalsf{}  &   &              &      &    & 5 & 34 & 58 & 11 & 1 & 8 & 1 & 1 & 28 & 11 & 58 & 7 & 8 & 19  \\     
    
    \vspace{-0.8em} \\
    \bottomrule
    \end{tabular}}
        \begin{tablenotes}\footnotesize
            \item[] (a) The benchmark data sets are specified as \textit{(x)f/(y)mol} where \textit{x} is the number of frequencies and \textit{y} the number of molecules considered in the data set.
            \item[] (b) \textit{Unc.} and \textit{Stand. Dev.} stand for uncertainty and standard deviation, respectively. \textit{None} is used when no statistical figure is reported.
            \item[] (c) \textit{LS} = least-squares, \textit{RS} = reduced scale factor optimisation model defined by Alecu and co-workers \cite{10AlZhZh}, \textit{MM} = MAD minimisation.
            \item[] (d) \textit{SE} = semiempirical, \textit{HF} = Hartree-Fock, \textit{H} = Hybrid, \textit{DH} = Double-hybrid, \textit{P-MP2} = Post-MP2.
            \item[] (e) Publications considering all vibrational frequencies in the benchmark dataset to derive the scaling factors.
            \item[] (f) Only some of the scaling factors reported in this reference have RMSE.
            \item[] (g) Publications focusing on one specific vibrational mode (e.g., C-O stretches) to derive the scaling factors. 
            \item[] (h) The values for the basis sets and levels of theory do not correspond to the total sum for each column. Instead, these values represent the count of distinct basis sets and levels of theory for each category.
        \end{tablenotes}
    \scalebox{0.755}{
    \begin{tabular}{lccl}
    \\
    Benchmark Set & Max atoms & Max nH atoms & Description \\
    \midrule
    3939f/358mol  & 20       & 6          & Molecules available in the CCCBDB database (atoms with atomic number less than 36 (Kr)). \\
    922f/90mol    & 18       & 8          & Small organic molecules containing second-, third- and fourth-row elements. \\
    1303f/65mol   & 16       & 10         & Variety of organic and inorganic molecules containing H, C, N, O and S.  \\
    510f/42mol    & 12       & 6          & Organic molecules containing second-row elements (i.e. Li-Ne). \\
    1069f/143mol  & 10       & 4          & Molecules and radical species containing second- and third-row elements. \\
    1064f/122mol  & 10       & 4          & Common organic and inorganic molecules containing second- and third-row elements. \\
    1033f/117mol  & 10       & 4          & Common organic and inorganic molecules containing second- and third-row elements. \\
    950f/125mol   & 10       & 4          & Organic molecules containing second-row elements. \\
    900f/111mol   & 10       & 4          & Common organic and inorganic molecules containing second- and third-row elements. \\
    486f/83mol    & 10       & 4          & Common organic molecules containing second-row elements.  \\
    824f/157mol   & 8        & 7          & Common organic and inorganic molecules containing second- and third-row elements. \\
    312f/63mol    & 8        & 4          & Molecules containing second- and third-row elements.  \\
    119f/30mol    & 6        & 3          & Molecules containing second- and third-row elements.  \\
    99f/26mol     & 6        & 3          & Molecules and radical species containing second-row elements. \\
    50f/15mol     & 4        & 3          & Molecules containing second-row elements.  \\
    18f/18mol     & 2        & 2          & Diatomic molecules and radical species from second-row elements. \\
     & & & \\
    C$\equiv$Of/31mol   & 51       & 28         & C$\equiv$O stretch frequencies present in 31 metal carbonyl complexes. \\
    COf/20mol     & 17       & 17         & C-O stretch frequencies present in 20 metal carbonyl complexes. \\
    COf/40mol     & 9        & 9          & C-O stretch frequencies present in 41 transition metal complexes. \\
    
    \bottomrule
    \end{tabular}}

\end{table*}

Based on a thorough literature review, summarised in \Cref{tab:lit_review}, we have compiled \totalsf{} uniform scaling factors derived for 141 levels of theory and 109 basis sets. 

We can order the different levels of theory in order of increasing computational accuracy (and usually cost) into nine different classes: \begin{compactenum}
    \item Semiempirical methods;
    \item Hartree-Fock (HF);
    \item Local Spin Density Approximation (LSDA);
    \item Generalised Gradient Approximation (GGA);
    \item meta Generalised Gradient Approximation (mGGA);
    \item Hybrid functionals;
    \item MP2;
    \item Double-hybrid functionals; and
    \item Post-MP2 methods (which includes methods such as CCSD(T) and QCISD)
\end{compactenum}
thus approximately merging one axis of the traditional Pople diagram for wave function methods \cite{65Po} with the Jacob's ladder for DFT procedures \cite{19GoMe}.

Similarly, we can also classify the basis sets in terms of their increasing zeta-quality (i.e., single-zeta, double-zeta, triple-zeta, etc.).

The data compiled here includes scaling factors derived for different spectral regions (i.e., global, low- and high-frequency ranges) with their respective statistical figures, as well as the optimisation methods and benchmark datasets used in their calculations. We report this information in \Cref{tab:lit_review} with references to the original publications, and the count of the aforementioned level of theory and basis set classes. The last row in \Cref{tab:lit_review} (\textit{Total}) corresponds to the total number of distinct entries for each basis set zeta-quality and level of theory class (note that these numbers do not correspond to the total sum across each column due to multiple use of several levels of theory and basis sets). 

\Cref{tab:lit_review} shows that most benchmark datasets share similar chemical compositions, differing primarily in the number of vibrational frequencies and molecules considered. %The least-squares method and RMSE are used in most publications over alternative optimisation methods and statistical figures. 

Focusing on the level of theory and basis set classes, \Cref{tab:lit_review} shows that there has been a somewhat arbitrary rather than systematic generation of scaling factors. Most scaling factors are from hybrid and GGA functionals, and double and triple-zeta basis sets, which can be attributed to the popularity and cost-effectiveness of these levels of theory and basis sets.

\begin{figure}[h]
    \centering
    \includegraphics[width=0.48\textwidth]{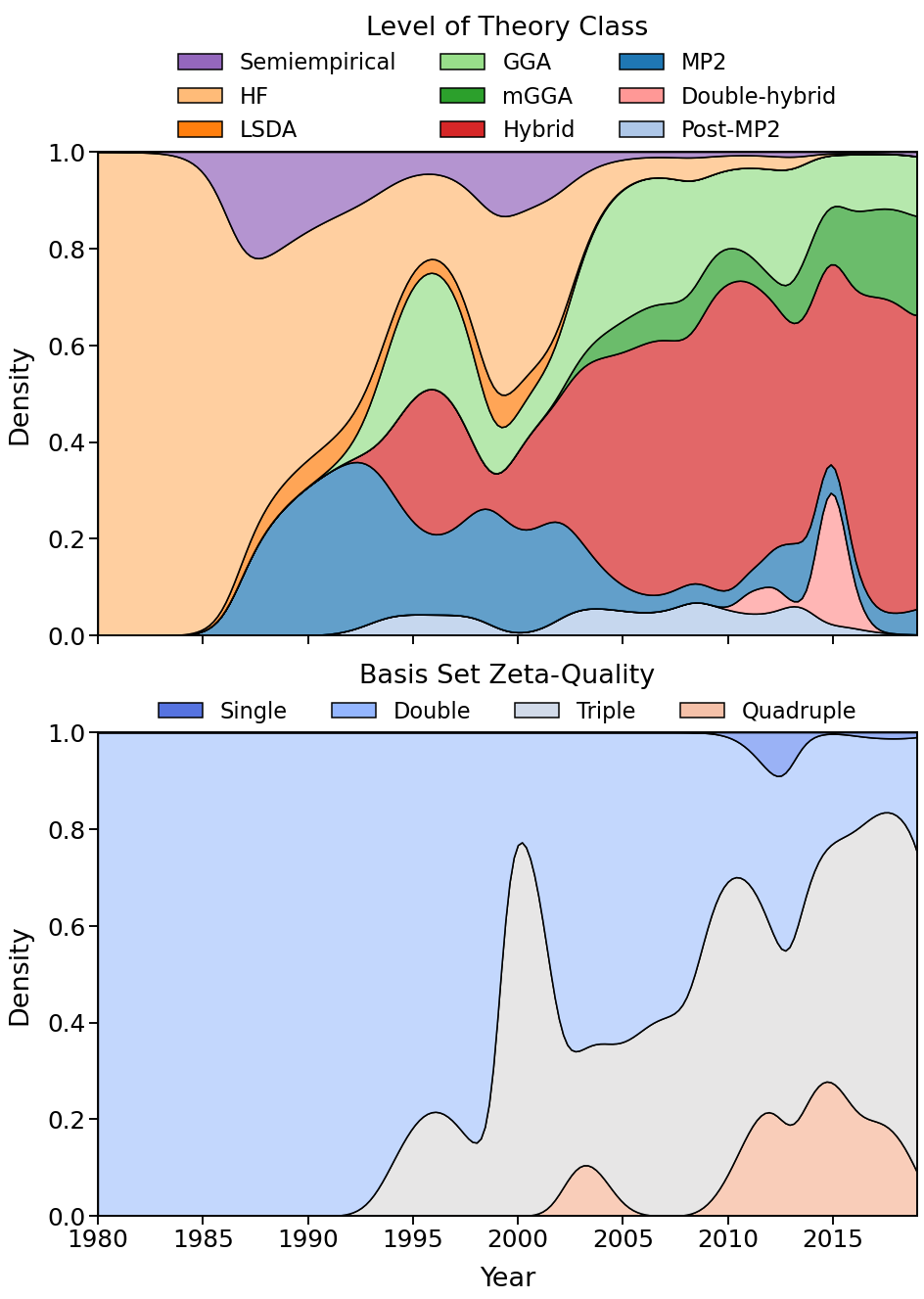}
    \caption{Density distribution of the different level of theory classes (top) and basis set zeta-qualities (bottom) used in the calculation of uniform scaling factors, as a function of the year of their publication.}
    \label{fig:density_dist}
\end{figure}

In \Cref{fig:density_dist} we present the density distribution of the compiled scaling factor data as a function of the year of their publication. The top figure shows that only Hartree-Fock scaling factors were produced between 1980 -- 1985, highlighting the first vibrational frequency calculations using computational quantum chemistry. A greater  heterogeneity in the level of theory  is observed between 1985 and 1995 as semiempirical, MP2 and pioneering density functional methods were  implemented into quantum chemistry engines. Since 1995, GGA and hybrid density functional approximations (DFAs) have dominated, with HF now largely eliminated, aligning with the large counts presented in \Cref{tab:lit_review} for the GGA and hybrid functional classes. mGGAs and double-hybrid functionals are more recent inclusions. Scaling factors for post-MP2 methods are only rarely calculated due to their large computational cost. 

With regards to the basis set zeta-quality, the bottom side of \Cref{fig:density_dist} confirms the extensive use of double- and triple-zeta basis sets in scaling factor calculations, with more recent studies providing more triple-zeta bases scaling factors.  Interestingly, we can also observe the contemporary usage of single-zeta basis sets in scaling factor calculations. Though somewhat surprising, this can be attributed to the development of cost-effective methodologies that rely on small basis sets choices to provide a quantum-mechanical treatment of medium- to large-sized systems \cite{13KaChRa, 16ChKaKa, 16BlBaBi}.

% improved computational power that now allows quantum-mechanical treatment of vibrations in medium- to larger-sized systems.

% very large systems like small proteins and polymers. \alert{Can we have a citation for the vibrational calculations in proteins and polymers -thanks!}

\begin{table}[h]
\centering
    \caption{Count of the different scaling factor types (i.e., global, low- and high-frequency) and statistical figures reported (in \cm{}). The \textit{None} column counts the number of scaling factors reported without any statistical figure whatsoever.}
    \label{tab:sf_stats}
    \scalebox{0.95}{
    \begin{tabular}{lcccccc}
        \toprule
        Scaling Factor Type & \mc{5}{c}{Statistical Figure} & Total \\
        \cmidrule(r){2-6}
         & \mc{1}{c}{RMSE} & \mc{1}{c}{MAD} & \mc{1}{c}{Uncert.} & \mc{1}{c}{Stand. Dev.} & \mc{1}{c}{None} & \\
         
        \midrule
        \vspace{-0.8em} \\
         
        Global         & 967  & 12 & 94 & 6  & 223 & 1302 \\ 
        Low Frequency  & 400  & - & -  & 6  & 31  & 437  \\
        High Frequency & 193  & - & -  & 6  & -   & 199  \\
        
        \vspace{-0.8em} \\
        \bottomrule
    \end{tabular}}
\end{table}

In \Cref{tab:sf_stats} we present the count for the different scaling factor types (i.e., global, low- and high-frequency) along with the four most common statistical reported; root-mean-square error (RMSE), mean absolute deviation (MAD), uncertainty (Uncert.) and standard deviation (Stand. Dev.). 
Global scaling factors are most common, especially from large compilations such as \citet{15KeBrMa}, \citet{17KaChNe} and \citet{10AlZhZh}, with low-frequency considerably more common than high-frequency scaling factors.

RMSE was by far the dominant statistical figure used and will be the figure considered for the rest of this manuscript. Interestingly, there are a significant number of scaling factors for which no statistical figure was available (listed in column \textit{"None"}). This is mostly attributed to those scaling factors derived following the procedure proposed by Alecu and co-workers \cite{10AlZhZh}, where the scaling factors for fundamental frequencies are calculated through a set of proportionality equations, rather than pursuing the minimisation of any statistical error.

All the scaling factor data compiled here are available in the supplemental material or upon request to the authors. The data includes the different statistical figures, benchmark data sets, optimisation method and references to the original publications. We note that there are two other repositories of uniform scaling factor data: the Computational Chemistry Comparison and Benchmark DataBase (CCCBDB) \footnote{https://cccbdb.nist.gov/vibscalejust.asp, date of access}, where the scaling factors are sorted in terms of levels of theory and basis sets, together with the uncertainties in their values and the list of molecules used in their calculations; and the Database of Frequency Scale Factors for Electronic Model Chemistries maintained by Kanchanakungwankul and co-workers \footnote{https://comp.chem.umn.edu/freqscale/, date of access}, which also stores scaling factors for zero-point energies and true harmonic frequencies. 

\section{History of Uniform Scaling Factors}
\label{sec:history}

Here, we detail the key development in scaling factors historically, focusing on highlighting changes in methodologies and scope.

\subsection{The Beginning of Scaling Factors}
\label{subsec:beginning}

Scaling factors arose to enable affordable quantum chemistry calculations of harmonic vibrational frequencies to be used to predict reasonable experimentally relevant fundamental frequencies, despite the calculation's neglect of anharmonic effects. 

In 1981, \citet{81PoScKr} pioneered this approach by noting that, for a data set covering 486 experimental fundamental bands, harmonic frequencies calculated at the HF/3-21G level were about 12\% larger than the experimental fundamental frequencies. Based on least-squares fitting between experimental fundamental frequencies and calculated harmonic frequencies, the authors recommended applying a multiplicative scaling factor of 0.89 to predict fundamental frequencies using harmonic HF/3-21G calculations. 

Soon after, different scaling factors were derived for the HF/6-31G* \cite{85DeMc, 89PoHeFo} and MP2/6-31G* \cite{82HoLeHe, 85DeMc} model chemistries, which were later improved by \citet{93PoScWo} using a larger benchmark data set with 1066 experimental vibrational frequencies from 122 molecules (1064f/122mol set in \Cref{tab:lit_review}). This data set remains today as the most popular benchmark choice for scaling factor calculations, and multiple sub-sets of the original data have also been used in several scaling factor studies. Scaling factors of 0.8929 and 0.9427 were recommended for HF/6-31G* and MP2/6-31G* with reported RMSE of 47 and 61\,\cm{} respectively.
  % to minimise some statistical deviation  between the predicted fundamental frequency $\lambda \omega_$

The simple approach adopted by Pople -- uniform multiplicative scaling of harmonic frequencies -- would, long-term, prove to be the most enduring. 

\subsection{Increasing Model Chemistry Diversity}
\label{subsec:dft}

With the advantages in computational efficiency delivered by the dawn of density functional theory, scaling factors were naturally derived for these new approaches.  Most influentially, in 1996, \citet{96ScRa} calculated scaling factors for fundamental frequencies and a variety of chemically-relevant properties at 19 different model chemistry choices. These choices included semiempirical, HF, MP2, QCISD and several DFT methods, together with multiple Pople-style basis sets. They found that all DFT-based methods considered represented good candidates for the calculation of vibrational frequencies, as demonstrated by their low RMSE and scaling factors closer to unity. Furthermore, they found the integration grid had negligible effects on both the scaling factor and RMSE values.% they investigated the effect of the integration grid in the calculations finding a negligible effect on both the scaling factor and RMSE values when increasing the grid size in the calculations (the recommended grid was a pruned version of a (50,194) grid). 
With more than 7000 citations, this publication remains as one of the most popular scaling factor studies performed. The popularity of B3LYP/6-31G* in vibrational frequency calculations can be traced predominantly to this paper's recommendations. 

Generation of scaling factors has continued in line with development of new quantum chemistry levels of theory. New density functional theories have of course been the dominant innovations; the large 2007 study by \citet{07MeMoRa} is particularly notable and found outstanding performance at the hybrid density functional level. In the last five years, scaling factors have also been computed for double-hybrid \cite{16ChRa} and dispersion-corrected functionals \cite{19Ha}. We note that generation of scaling factors for fundamental frequencies has also kept pace with ongoing development of new high-level wavefunction \cite{11DoBe,14FrToHa}, fast semi-empirical \cite{04WiMo, 07FeHoKo}, and low-cost procedures \cite{17Cha}.

Similarly, basis set diversity for scaling factors has increased substantially from the original Pople and then Dunning families, which are not recommended for modern DFT calculations \cite{nagy2017basis,jensen2013atomic}, to more contemporary basis sets. Indeed, we can nowadays find scaling factors for model chemistries including a wide variety of basis set families: Dunning correlation-consistent (cc-pV$n$Z) \cite{04SiBoGu, 04WiMo, 10AlZhZh, 10TeMeCo, 11LaBoHa, 11DoBe, 14FrToHa, 15KeBrMa, 16ChRa, 17Cha, 17KaChNe}, Ahlrichs-Karlsruhe (def2-$n$) \cite{10AlZhZh, 15AsDeBr, 15KeBrMa, 16ChRa}, Jensen (pc-$n$) \cite{12LaCaWi, 15KeBrMa}, LANL \cite{15AsDeBr, 17Cha}, Sadlej \cite{01HaVeSc, 17KaChNe},   Barone \cite{15KeBrMa}, Huzinaga \cite{17Cha} Jorge ($n$ZP) \cite{08AnGoJo}, Petersson \cite{15KeBrMa}, Sapporo \cite{17KaChNe},   Truhlar \cite{10AlZhZh}, and STO-$n$G \cite{17Cha}.

Individual papers do generally compare the scaling factors and RMSE for the model chemistries they consider, yet the scope of any one paper yet published is not sufficiently diverse in level of theory and basis set to be considered comprehensive.  One of the primary purposes of this review is to unify and evaluate the existing data to compare model chemistries. 

\subsection{Data Science Considerations}

\label{subsec:stats}

\textbf{Separation of Training and Test Set:} Traditionally, scaling factor papers have unfortunately only reported performance for a training dataset rather than an independent test dataset, unsatisfactory from a data science perspective. We praise the recent work of \citet{21UnNaOz} in dividing their benchmark set of molecules into a training (82.2\% of the original data) and a test (remaining 17.8\%) set and strongly recommend this approach for all future studies. 
=

\textbf{Significant Figures:} In their study of the uncertainty of scaling factors, \citet{05IrRuRa} found that scaling factors are only accurate up to the second significant figure, despite being commonly reported to four or more significant figures. %\alert{Juan: Did 05IrRuRa use the same approach as 10TeMeCo?} \alertjuan{Yeah! 10TeMeCo used the same methodology as 05IrRuRa to calculate their scaling factors.} 
\citet{10TeMeCo} found similar results with the X3LYP functional by considering scaling factors for subsets of the benchmark dataset.

\subsection{Alternative Scaling Factor Approaches}

\textbf{Low- and High-Frequency Scaling Factors:} As early as 1996, \citet{96ScRa} recognised that global scaling factors derived using standard least-squares approaches were more sensitive to high-frequency than low-frequency vibrations. This limitation can result in a less accurate interpretation of the fingerprint region (approximately below 1000\,\cm{}) or in poor prediciton of thermochemical properties such as the thermal contributions to enthalpy and entropy, which are heavily dominated by low-frequency vibrations. To address this issue, \citet{96ScRa} proposed an inverse frequency scaling factor calculated as $\lambda_\textrm{low-freq} =  \sum_{i}^{N} \left(\frac{1}{\omega_{i}}\right)^{2} / \sum_{i}^{N}\frac{1}{\omega_{i} \nu_{i}}$, which is more weighted towards low-frequency than high-frequency vibrations. 
Though initially uncertain, the best choice of threshold between low- and high-frequency ranges was considered in 2004 by \citet{04SiBoGu} who concluded an optimal cutoff of 1000\,\cm{}. 

Low-frequency scaling factors have subsequently been calculated in multiple studies \cite{01HaVeSc, 04SiBoGu, 05AnUv, 07MeMoRa, 08AnGoJo, 11LaBoHa, 12LaCaWi, 16ChRa, 17Cha}, attesting to the likely success of this method. We note that  direct quantification of the advantages of low-frequency factors compared to global scaling factors have not been explicitly   considered in depth.

High-frequency scaling factor studies \cite{01HaVeSc,04SiBoGu,11LaBoHa, 12LaCaWi} are  rarer likely because both global and high-frequency scaling factors share a very similar values and performance.

\textbf{Band-specific Scaling:} By constraining the benchmark dataset to only consider specific types of frequencies, higher accuracy in the scaling factor's performance can be obtained. Examples of this approach are the works of \citet{03YuSrSc,08SiMaEh} and \citet{15AsDeBr} who constrained their benchmark data sets to only consider the C-O stretch frequency in multiple organometallic complexes. Indeed, the RMSEs obtained through this method are typically around 12\,\cm{}, but can be as small as 6\,\cm{} with appropriate model chemistry choices. 

This approach is certainly a viable way to improve accuracy of band-specific predictions, but in practical terms we expect only narrow areas of applicability.

\textbf{Non-multiplicative Scaling: }More complex approaches have used linear \cite{00YoEhMa, 00Al, 02YoTaOk, 02AlRa, 02PaIzGi, 19Pa} and polynomial \cite{19Pa} equations to derive scaling factors for harmonic frequencies. These procedures inherently lead to further improvements in the scaled frequencies as the scaling changes depending on the different spectral regions, i.e., the same scaling factor value is not applied to all frequencies. RMSE values ranging between 10 -- 20\,\cm{} have been found through application of these methods \cite{00Al, 02AlRa, 02AlNuGi, 02PaIzGi, 05PaGiNu, 18AlRaSi, 18AlKaAf}. However, the parameters obtained in these approaches are highly dependent upon the molecules considered, limiting their broad applicability.

\textbf{Force-constant Scaling:} Mode-specific scaling of force constants have been investigated extensively, as reviewed by \citet{19BaBo} and \citet{19Pa}. Improved prediction accuracy can be obtained (e.g., RMSE below 15\,\cm{}); however, their complexity and reliance on available benchmark data for the considered mode limits routine applications in quantum chemistry calculations. 

% \textbf{Quick Scaling Factors Generation:} Approaches to quickly calculate scaling factors for multiple model chemistries without computing statistical figures have also been proposed in the literature. However, though appealing for their quickness, these approaches cannot provide any insight regarding the model chemistry performance, i.e., no overall error can be estimated for the calculated frequencies after scaling. An example of these approaches is the work by \citet{10AlZhZh} (later automated by \citet{17YuFiAl}), where a small benchmark dataset is used to derive scaling factors for fundamental frequencies and other properties. 

\section{Analysis of Existing Scaling Factors Data} 
\label{sec:SF_existing_data}

Scaling factor performance is assessed in this manuscript entirely through reported RMSE between experiment and calculated values. We also constrain our discussion to scaling factors derived using global benchmark data sets instead of band-specific data sets (see \Cref{tab:lit_review}).
%Whilst acknowledging the diversity of statistical figures reported along with the scaling factor, we constrained ourselves to pursue further analysis on scaling factors reported along with RMSE only, given their popularity and diversity in terms of levels of theory and basis sets (see \Cref{tab:lit_review} and \Cref{tab:sf_stats}). Along these lines, we would also like to make the reader aware that in the following sections the term `scaling factor performance' will be used to refer to how close the scaled harmonic frequencies are to experimental frequencies.

% We understand that the different statistical figures reported in \Cref{tab:sf_stats} provide useful information regarding the scaling factor's performance. However, due to their popularity and diversity in terms of levels of theory and basis sets, further discussions and analysis in this manuscript are constrained to scaling factors reported along with RMSE. 

\begin{figure}[h]
    \centering
    \includegraphics[width=0.48\textwidth]{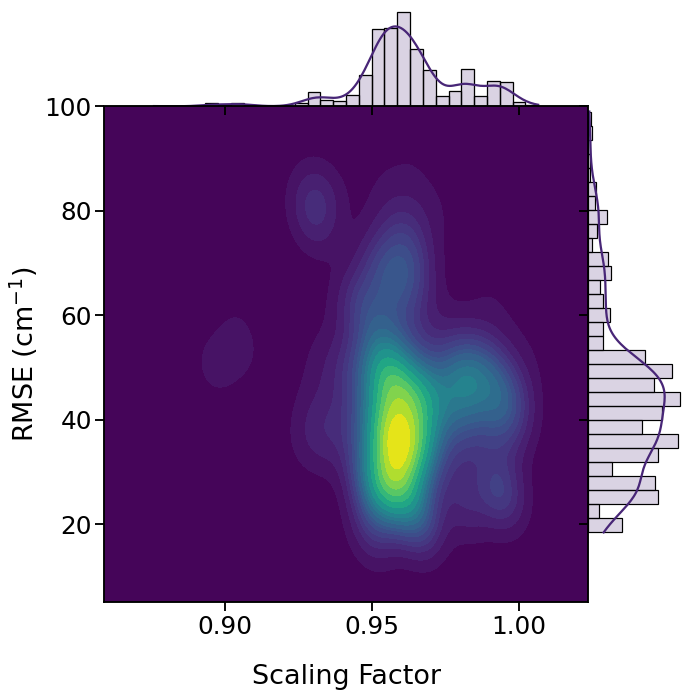}
    \caption{Comparison between global scaling factors and their associated RMSE (\cm{}). The histograms on the top and right sides of the figure present the distributions of the scaling factor values and RMSE, respectively. The colour transition in the middle figure shows the correlation between the scaling factor values and the RMSE, with the correlation increasing from purple to yellow. For readability, only scaling factors ranging between 0.80 and 1.15, and RMSE between 0 to 100\,\cm{} have been considered.}
    \label{fig:rms_vs_sf}
\end{figure}

\subsection{Global Scaling Factors}

\Cref{fig:rms_vs_sf} details global scaling factors and their associated RMSE. The figure shows no clear relationship between the two variables, but demonstrates that most scaling factors are around 0.96 while RMSE are usually between 20 -- 50\,\cm{}. The value of the scaling factor thus cannot be used as a proxy for model chemistry performance. 

Most methods do not perform significantly better than the 25\,\cm{} lower bound predicted by \citet{15KeBrMa} using the 119f/30mol dataset in \Cref{tab:lit_review}. Lower RMSE are probably attributable to cancellation of errors or differences in the benchmark dataset. % or reduced diversity of the benchmark dataset (e.g., only considering C-O stretch frequencies). 

% This observation indicates that the scaling factor value \textit{per se} cannot be used as a proxy of the model chemistry performance in vibrational frequency calculations as, for different model chemistries, similar scaling factors values are reported along with divers RMSE. We can corroborate this by noting the larger spread in the RMSE histogram (left) when compared to that for the scaling factor values (top). On the other hand, the RMSE distribution shows that, for a given model chemistry, the lowest RMSE is reached at $\sim$ 20 \cm{}, defining a bottom boundary for the global scaling factor's performance. These results align with the findings of Kesharwani and co-workers who noted that scaled harmonic frequencies are expected to be at least 25 \cm{} off of their experimental counterparts \cite{15KeBrMa}. Lower RMSE may be achieved for some particular model chemistries, but this might be most likely due to cancellation of errors throughout the calculations.

\begin{figure}[h]
    \centering
    \includegraphics[width=0.48\textwidth]{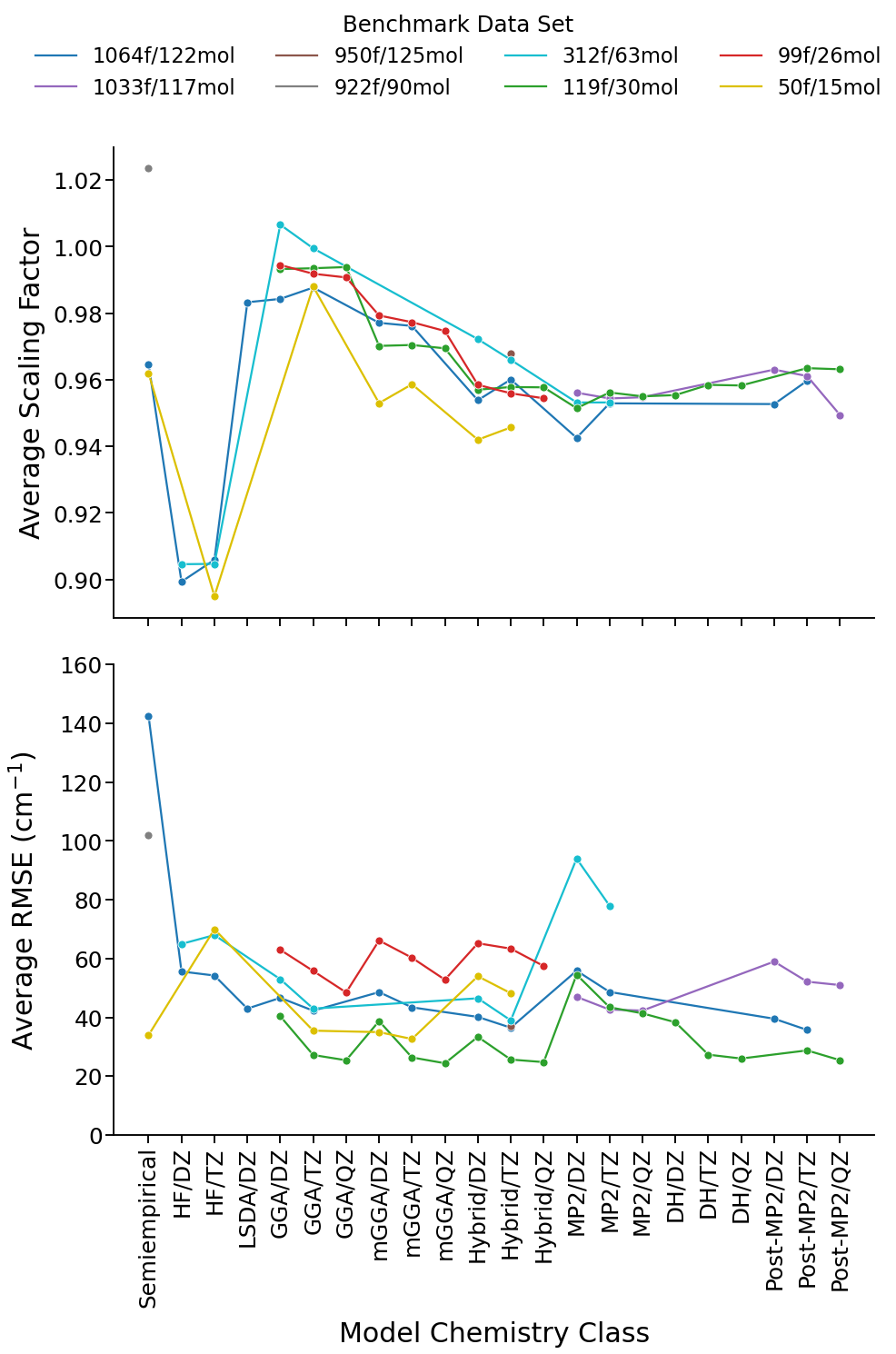}
    \caption{Average scaling factor values and RMSE as a function of the different model chemistry classes (i.e., level of theory class/basis set zeta-quality). The data points are coloured according with the benchmark data set used in the calculation of the scaling factors. Data points in the figure are joined only to aid readability.}
    \label{fig:LTClass_average}
\end{figure}

\Cref{fig:LTClass_average} presents the average of both the scaling factor and RMSE across different level of theory classes and basis set zeta-qualities (aka model chemistry class) over the full set of compiled data. The model chemistry classes are ordered approximately in order of increasing computational time from left to right, keeping level of theory classes together.

%We can investigate further the larger variation in the RMSE by considering three different aspects: the basis set zeta-quality, the level of theory class and the molecular types used to derive the scaling factor (i.e., the benchmark data sets). In \Cref{fig:LTClass_average} we , over the benchmark data sets presented in \Cref{tab:lit_review} (note that only data sets used to derive scaling factors along with RMSE are included in this analysis).

%The top figure shows that the benchmark data set choice has little effect in the scaling factor value, especially when mGGA/DZ or more robust model chemistry classes are used. Below the mGGA/DZ point, we can observe more variability in the scaling factor value when changing the benchmark set (see for example the GGA/DZ or semiempirical classes). 

As the computational cost of the model chemistry classes increases, the top sub-figure in \Cref{fig:LTClass_average} shows that the scaling factor converges by about Hybrid/DZ level to roughly 0.96 with no systematic improvement in RMSE at more computationally expensive levels of theory, though some improvement is observed with larger basis sets. These trends indicate that errors from inadequate treatment of anharmonicity dominate over electron correlation and incomplete basis set errors even at the Hybrid/DZ-TZ levels. These results are critical, suggesting that higher levels of theory are unnecessary and could be avoid when predicting fundamental frequencies through multiplicative scaling of calculated harmonic frequencies.

\Cref{fig:LTClass_average} also shows that the chosen benchmark data set only minimally impacts the scaling factor but conversely has a major impact on the RMSE. For example, the 99f/26mol (red) benchmark set exhibits the largest average RMSE; this can be attributed to the high proportion of radical species in the calculations which cause known problems for DFT computations  \cite{09BrLiGi,12TeAr,15YuTr,16YuLiTr,18Hs}. 

As our data was compiled from the literature rather than being regenerated, we note there is non-uniform treatment of different methodologies. In particular, there are 109 model chemistries for which at least two scaling factors have been proposed in the literature (in some cases, up to seven scaling factors are available for the same model chemistry). In \Cref{tab:repeated_mc} we only present six model chemistries with multiple scaling factors reported along with their corresponding statistical figures and benchmark data sets used. In all cases, we can observe that very similar scaling factors have been found, supporting our previous finding that the benchmark data set has minimal effect on the computed scaling factor.  The RMSE, on the other hand, depend significantly on the benchmark dataset selected; the RMSE values in the M06 family are much larger for the 119f/30mol dataset than the 99f/26mol dataset due to the large number of radical species present in the former dataset. %In fact, for several of the model chemistries with multiple scaling factor that have used the 99f/26mol benchmark set, the standard deviations in their RMSE range between 15 -- 35\,\cm{}, as visually displayed in figure S2 in the supplementary material. 

\begin{table}[h]
\centering
    \caption{Scaling factors and statistical figures (in \cm{}) for some model chemistries with multiple entries reported in the literature.}
    \label{tab:repeated_mc}
    \scalebox{0.8}{
    \begin{tabular}{lcllll}
        \toprule
        Model Chemistry                 & Scaling Factor & Stat   & Stat Type & Data Set     & Ref      \\
        \midrule
        \vspace{-0.8em} \\
        
        \multirow{7}{*}{B3LYP/6-31G(d)} & 0.9614         & 34     & RMSE       & 1064f/122mol & \cite{96ScRa}   \\
                                        & 0.9613         & 34     & RMSE       & 1064f/122mol & \cite{96Wo}     \\
                                        & 0.9613         & 34     & RMSE       & 1064f/122mol & \cite{07MeMoRa} \\
                                        & 0.9636         & 36.18  & RMSE       & 119f/30mol   & \cite{15KeBrMa} \\
                                        & 0.9654         & 0.028  & Uncert.   & 1069f/143mol & \cite{10TeMeCo} \\
                                        & 0.9594         & 0.02   & Uncert.   & 3939f/358mol & \cite{05IrRuRa} \\
                                        & 0.9520         & -   & -      & 50f/15mol    & \cite{10AlZhZh} \\
        \vspace{-0.4em} \\
        \multirow{6}{*}{BLYP/6-31G(d)}  & 0.9940         & 45     & RMSE       & 1064f/122mol & \cite{07MeMoRa} \\
                                        & 0.9945         & 45     & RMSE       & 1064f/122mol & \cite{96ScRa}   \\
                                        & 0.9940         & 45     & RMSE       & 1064f/122mol & \cite{96Wo}     \\
                                        & 0.9903         & 0.0253 & Uncert.   & 3939f/358mol & \cite{05IrRuRa} \\
                                        & 0.9830         & -   & -      & 50f/15mol    & \cite{10AlZhZh} \\
                                        & 0.9940         & -   & -      & 1064f/122mol & \cite{17Cha}    \\
        \vspace{-0.4em} \\
        \multirow{7}{*}{HF/6-31G(d)}    & 0.8953         & 50     & RMSE       & 1064f/122mol & \cite{96ScRa}   \\
                                        & 0.8953         & 50     & RMSE       & 1064f/122mol & \cite{93PoScWo} \\
                                        & 0.8953         & 50     & RMSE       & 1064f/122mol & \cite{07MeMoRa} \\
                                        & 0.8981         & 0.0334 & Uncert.   & 1069f/143mol & \cite{10TeMeCo} \\
                                        & 0.8982         & 0.023  & Uncert.   & 3939f/358mol & \cite{05IrRuRa} \\
                                        & 0.8850         & -   & -      & 50f/15mol    & \cite{10AlZhZh} \\
                                        & 0.8900         & -   & -      & 486f/83mol   & \cite{81PoScKr} \\
        \vspace{-0.4em} \\
        \multirow{2}{*}{M06-L/cc-pVDZ}  & 0.9630         & 40.34     & RMSE       & 119f/30mol & \cite{15KeBrMa} \\
                                        & 0.9649         & 84.00     & RMSE       & 99f/26mol & \cite{17KaChNe} \\
        \vspace{-0.4em} \\
        \multirow{2}{*}{M06-2X/cc-pVDZ} & 0.9543         & 35.42     & RMSE       & 119f/30mol & \cite{15KeBrMa} \\
                                        & 0.9510         & 72.00     & RMSE       & 99f/26mol & \cite{17KaChNe} \\
        \vspace{-0.4em} \\
        \multirow{2}{*}{M06/cc-pVDZ}    & 0.9620         & 75.00     & RMSE       & 119f/30mol & \cite{15KeBrMa} \\
                                        & 0.9655         & 42.04     & RMSE       & 99f/26mol & \cite{17KaChNe} \\
        \vspace{-0.8em} \\                                
        \bottomrule
    \end{tabular}}
\end{table}

\subsection{Low- and High-frequency Scaling Factors}

Here, we focus on low- and high-frequency scaling factors, noting the relative sparsity of the data will limit the generalisability of our conclusions. 

\begin{figure}[h]
    \centering
    \includegraphics[width=0.48\textwidth]{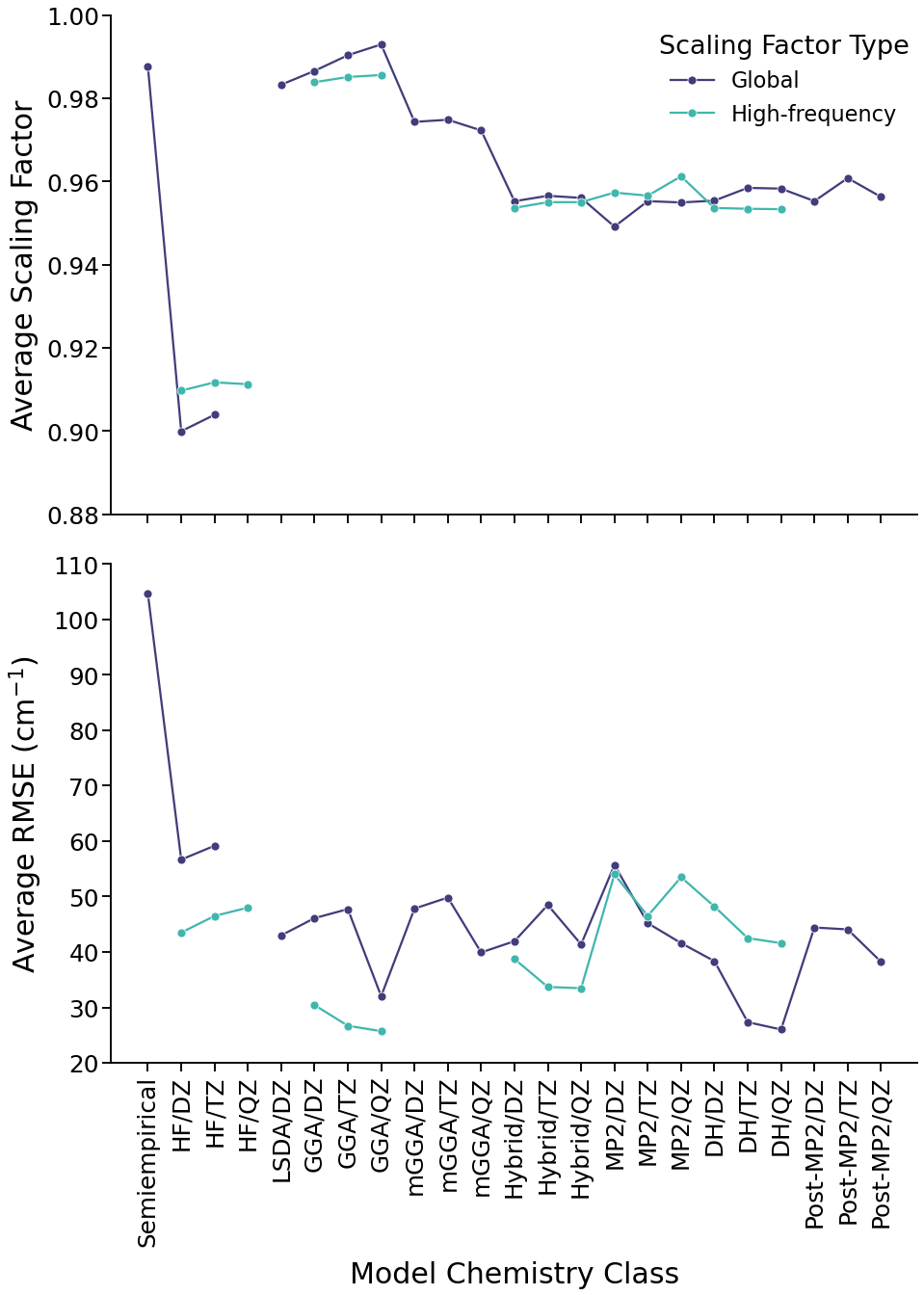}
    \caption{Average scaling factor values and RMSE for the global and high-frequency ranges as a function of the model chemistry class. Data points in the figure are joined only to aid readability.}
    \label{fig:global_high_avg}
\end{figure}

\Cref{fig:global_high_avg} demonstrates that the global and high-frequency scaling factors and RMSE are very similar, justifying the relative neglect of high-frequency-specific scaling factors in the literature (199/1463 compiled scaling factors).

\begin{figure}[h]
    \centering
    \includegraphics[width=0.48\textwidth]{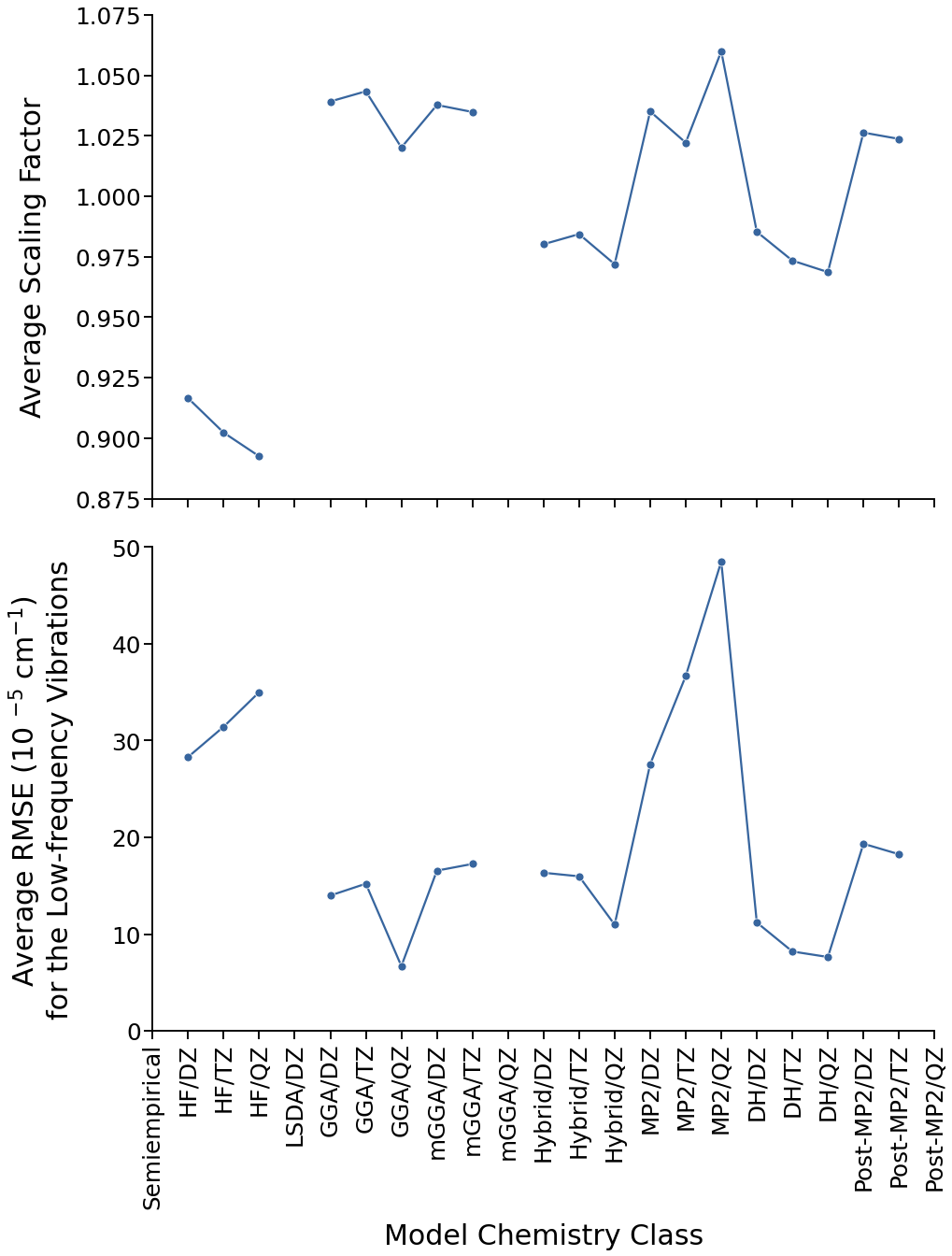}
    \caption{Average scaling factors and RMSE for the low-frequency range as a function of the model chemistry class. Data points in the figure are joined only to aid readability.}
    \label{fig:low_avg}
\end{figure}

\Cref{fig:low_avg}, on the other hand, shows that low-frequency scaling factors and RMSE show little convergence as a function of model chemistry class. We attribute this behaviour to the more complex nature of low-frequency vibrations, where large parts of the molecule collectively vibrate instead of localised functional groups.  It is likely that the harmonic approximation is insufficiently accurate for reliable scaling factors in this region given the abundance of large amplitude motions (LAMs), methyl rotations and torsional states.

\section{Best Model Chemistry According to Existing Data}
\label{sec:best_mc}

The primary motivation for this work was to understand which model chemistry choices represent the best options for the calculation of vibrational frequencies. Ideally, RMSE could be used to compare the performance of different model chemistries and identify superior approaches. Current data, however, does not permit a thorough comparison for two main reasons. First, the diversity of benchmark datasets means RMSE computed in different studies cannot be fairly compared. Second, the RMSE are never reported for an independent test set of molecules, i.e. there is almost never a separation of the training dataset used to fit the scaling factor and the test dataset used to determine its performance. A notable and praiseworthy exception is the recent work of  \citet{21UnNaOz}; we strongly encourage this approach in future studies calculating and evaluating harmonic frequency scaling factors. 

It is therefore premature to make definitive conclusions on the best performing model chemistries. However, we can provide preliminary insights based on available data, especially given the dominance of the 1064f/122mol dataset. 

We have constrained detailed consideration to hybrid and double-hybrid functionals combined with double- and triple-zeta basis sets to balance adequate performance with modest computational time. We have also excluded data produced from the 99f/26mol benchmark data sets, due to the dominance of problematic molecular species. This filtered subset of data corresponds to 323 scaling factors produced with 38 levels of theory (35 hybrid and 3 double-hybrid) and 35 basis sets (12 double-zeta and 23 triple-zeta).

\begin{table}[h]
\centering
    \caption{Count, average values and standard deviations ($\sigma$ in \cm{}) for the hybrid and double-hybrid functionals, as well as double- and triple-zeta basis sets considered in the filtered subset of data.}
    \label{tab:avg_best_mc}
    \scalebox{0.8}{
    \begin{tabular}{lccccc}
        \toprule
                                            & Count & Average SF & $\sigma$ & Average RMSE (\cm{}) & $\sigma$ \\
        \midrule
        \vspace{-0.8em} \\
            \textit{Level of Theory Class}  &       &            &             &                     &          \\
            Hybrid                          & 35    & 0.9560     & 0.012       & 36.64               & 10.52      \\
            Double-hybrid                   & 3     & 0.9572     & 0.005       & 32.05               & 6.72      \\
                                            &                                &                               \\
            \textit{Basis Set Zeta-Quality} &       &            &             &                     &          \\
            Double                          & 12    & 0.9549     & 0.011       & 38.83               & 9.67      \\
            Triple                          & 23    & 0.9574     & 0.011       & 33.48               & 10.06      \\
        \vspace{-0.8em} \\                                
        \bottomrule
    \end{tabular}}
\end{table}

Summary information of our filtered dataset is presented in \Cref{tab:avg_best_mc}, with average and standard deviation for the scaling factors and RMSE for different level of theory classes and basis set zeta-qualities. Within our limited data, double hybrids slightly outperform hybrid density functionals and triple-zeta slightly outperform double-zeta basis sets with overlapping standard deviations.

\begin{figure*}[ht]
    \centering
    \includegraphics[width=1\textwidth]{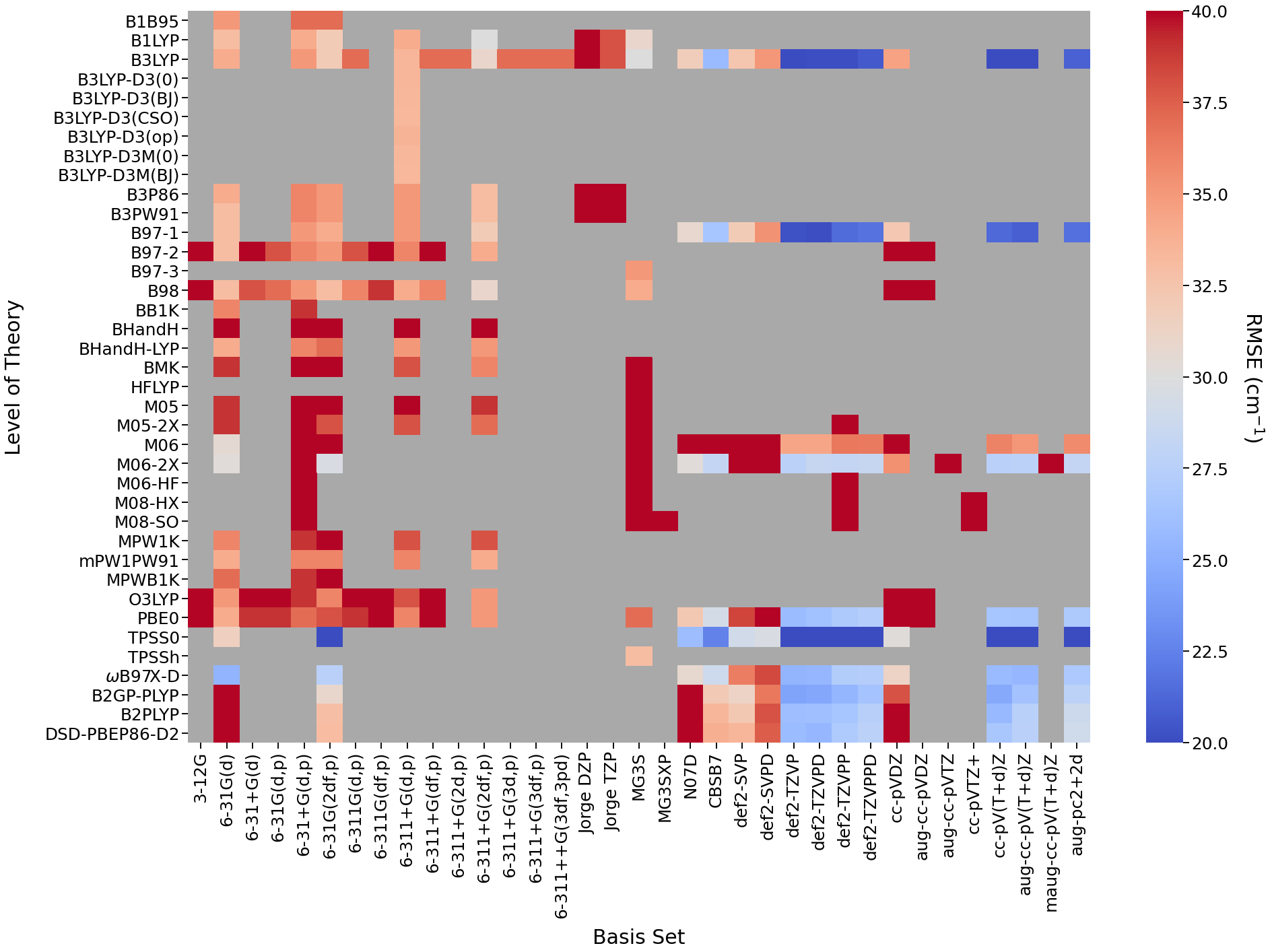}
    \caption{RMSE as a function of the different hybrid and double-hybrid functionals, and double- and triple-zeta basis sets considered in the filtered data. These results are taken from literature and not regenerated; note in particular that the benchmark set may be different for different datapoints (though results from the more challenging 99f/26mol dataset have been excluded). }
    \label{fig:heatmap}
\end{figure*}

\Cref{fig:heatmap} graphically shows the RMSE as a function of model chemistry choice, where blue and red represent strong and poor performance, respectively, and grey represents model chemistries for which no data is currently available.  We note the absence or sparsity of scaling factors for many more modern levels of theory and basis sets. 

There are multiple density functionals that delivery satisfactory performance (i.e., RMSE below 30\,\cm{} and close to the limit of anharmonicity error), notably $\omega$B97X-D, B3LYP, PBE0, DSD-PBEP86-D2, B2PLYP, B2GP-PLYP, TPSS0 and B97-1 functionals. Noting benchmarking results for accurate dipole moments \cite{18HaHe,20ZaMc} (and thus intensities), general-purpose DFT benchmarking results \cite{17GoHaBa,17MaHe} and the predicted maximimum accuracy of 25 \cm{} \cite{15KeBrMa}, we recommend B2PLYP as the best double-hybrid and $\omega$B97X-D as the best hybrid functional.  

Basis set choices seem to be more important; triple-zeta basis sets do significantly enhance performance over double-zeta basis sets in this detailed comparison, with the triple-zeta Alhrich def2-$n$ and Dunning cc-$n$ families significantly outperforming triple-zeta Pople-style basis sets.  \Cref{fig:heatmap} shows strong performance for aug-pc2+2d, aug-cc-pV(T+d)Z, cc-pV(T+d)Z, def2-TZVP, def2-TZVPD, def2-TZVPP, def2-TZVPPD. Based on ease of availability and the fact that Alhrich basis sets are optimised for DFT and are usually faster than the Dunning basis sets, we recommend def2-TZVP for frequency calculations. If accurate vibrational intensities (i.e., dipole moments) are also needed, diffuse functions should be included in the basis set \cite{20ZaMc} and therefore we recommend the def2-TZVPD basis set.

Given the ubiquity of B3LYP/6-31G* (also known as B3LYP/6-31G(d)), it is worth explicitly considering double-zeta basis set performance. \Cref{fig:heatmap} shows that there is far stronger coupling between level of theory and basis set in determining performance for double-zeta basis sets than triple-zeta basis sets and one must consider model chemistry. Considering the raw results, the most informative comparison is the relative performance  of $\omega$B97X-D/6-31G* (RMSE = 25 \cm{})  and B3LYP/6-31G* (RMSE = 36 \cm{}) using the same 119f/30mol benchmarking set by \cite{15KeBrMa}. The polarised double-zeta Pople basis sets deliver similar performance to other basis sets, unlike for the triple-zeta case. 

Our analysis thus far recommends three model chemistries in increasing order of computational cost tabulated in \Cref{tab:data_best_mc}. All model chemistries have negligible model chemistry errors compared to anharmonicity errors within our current data limitations.  Despite this, the increased robustness of the triple-zeta basis sets across different density functionals means we would usually prefer $\omega$B97X-D/def2-TZVP compared to $\omega$B97X-D/6-31G* if computational resources permit. If accurate intensities are desired, significant improvements should be observed using B2PLYP/def2-TZVPD (based on results of \citet{18HaHe,20ZaMc}.

% Our analysis thus far recommends three model chemistries in increasing order of computational cost: $\omega$B97X-D/6-31G* (SF = 0.9501, RMSE = 25 \cm{}), $\omega$B97X-D/def2-TZVP (SF = 0.9546, RMSE = 25 \cm{}), B2PLYP/def2-TZVPD (SF = 0.9636, RMSE = 26 \cm{}). All model chemistries have negligible model chemistry errors compared to anharmonicity errors within our current data limitations.  Despite this, the increased robustness of the triple-zeta basis sets across different density functionals means we would usually prefer $\omega$B97X-D/def2-TZVP compared to $\omega$B97X-D/6-31G* if computational resources permit. If accurate intensities are desired, significant improvements should be observed using B2PLYP/def2-TZVPD (based on results of \citet{18HaHe,20ZaMc}.

\begin{table}[ht]
\centering
    \caption{Scaling factors and RMSE for the recommended model chemistries. Data from \citet{15KeBrMa}.}
    \label{tab:data_best_mc}
    \begin{tabular}{lcc}
        \toprule
            Model Chemistry & Scaling Factor & RMSE (\cm{}) \\
            \cmidrule(r){1-3}
        \vspace{-0.8em} \\
            $\omega$B97X-D/6-31G*    & 0.9501 & 25.3 \\
            $\omega$B97X-D/def2-TZVP & 0.9546 & 25.4 \\
            B2PLYP/def2-TZVPD        & 0.9636 & 26.1 \\
        \vspace{-0.8em} \\                                
        \bottomrule
    \end{tabular}
\end{table}

\section{Conclusions and Future Directions}% Recommendations}
\label{sec:conclusions}
When predicting fundamental frequencies through multiplicative scaling of harmonic frequencies, there are three main sources of error; anharmonicity error, level of theory error, and basis set incompleteness error. The anharmoncity error limits the accuracy of this approach to a RMSE of 25\,\cm{} between predicted and experimental fundamental frequencies \cite{15KeBrMa}.

Existing data is sparse with evaluation challenges. Nevertheless, our compilation strongly supports the hypothesis that appropriately selected hybrid or double-hybrid functionals with double- or triple-zeta basis sets will reduce the level of theory and basis set incompleteness errors to sufficiently below the anharmonicity error as to be insignificant.  More computationally expensive levels of theory (e.g., CCSD(T) and quadrupole-zeta basis sets) are thus not recommended when predicting fundamental frequencies within the scaled harmonic approximation.

% Our primary recommendation for users predicting fundamental frequencies through multiplicative scaling of harmonic frequencies are that: 
% \begin{itemize}
%     \item density functional theory (not wavefunction methods) should always be used, with  hybrid or double-hybrid functionals  recommended (double hybrid functionals outperform MP2,  hybrid functionals outperform HF and improvements from using coupled-cluster methods cannot be realised because of the dominance of the anharmonicity error compared to model chemistry errors); 
%     \item triple-zeta basis sets are sufficiently accurate and quadrupole-zeta basis sets should be avoided as they offer no accuracy improvements. 
% \end{itemize}

Data sparsity and evaluation differences mean it is premature to provide definitive model chemistry recommendations, but existing data suggests that appropriate density functional and basis set choices are important to achieve optimal accuracy. $\omega$B97X-D/def2-TZVP is the recommended model chemistry for frequencies, with B2PLYP/def2-TZVPD recommended if accurate intensities are desirable and $\omega$B97X-D/6-31G* recommended if only double-zeta basis sets are computationally tractable.  We caution that (1) without separation of test and training data and with differing benchmark datasets, the RMSE on which performance is currently evaluated should be viewed as indicative only, and (2) many modern levels of theory (e.g. $\omega$B97X-V) and basis sets are not considered in our analysis due to the scarcity of scaling factors data currently available. %The $\omega$B97X-D, PBE0, DSD-PBEP86-D2, B2PLYP, B2GP-PLYP, TPSS0 and B97-1 density functionals all have satisfactory performance. For basis sets, def2-TZVP is recommended, with def2-TZVPD preferred if accurate intensities are desired. If only double-zeta basis sets are practical, def2-SVPD is preferred over Pople-style basis sets such as 6-31G(d). 

Given the widespread calculation of fundamental frequencies through scaled harmonic frequencies, broad benchmarking to identify high-performing model chemistries for this property within the hybrid/double-hybrid DFT and double/triple-zeta basis set classes is an urgent priority. For reliability, these studies should include separation of the training dataset used to compile the scaling factor and the test dataset used to evaluate its performance. 

The quick convergence of the scaling factor with model chemistry motivates speculation on whether a universal scaling factor might be preferable to a model-chemistry-specific value for all but the smallest model chemistries. Our results show the average scaling factor value across all model chemistry classes at or above the Hybrid/DZ level is 0.9560 with a small standard deviation of just 0.010; for comparison \citet{15KeBrMa} found a scaling factor of experimental fundamental frequencies to experimental harmonic frequencies of 0.9627(11). Noting that computed scaling factors are generally reliable only to 2 significant figures \cite{05IrRuRa}, 0.96 would be a useful universal scaling factor to test. %A universal scaling factor might therefore be preferable to a model-chemistry specific value for all but the most modest calculations, though further studies will of course be needed to confirm the viability of this approach. 

The implications of this quick convergence to a anharmonicity-dominated error should be considered further for important related contexts, e.g., thermodynamic composite method predictions (e.g., G$n$, W$n$ methods) and hybrid predictions of fundamental frequencies (with very accurate harmonic and cheap anharmonic calculations) \cite{10BiPaSc,14BaBiBl,18BiBlPu}. However, the use of more sophisticated anharmonic correction terms than a simple multiplicative factor could very well mean that the accuracy of the electronic structure methodology within the harmonic calculation becomes more critical than the specific (but extremely common) application which is the focus of this paper. 

Our results show that high-frequency specific scaling factors are unnecessary, as they perform very similarly to global scaling factors, but suggest low-frequency scaling factors could be useful. We note that low-frequency scaling factors have inconsistent convergence behaviour, i.e., there is no clear trend followed by the low-frequency scaling factor value when increasing the model chemistry class accuracy, likely due to the complexity of the underlying vibrations. Future work should also explicitly consider quantification of the comparative performance of low-frequency scaling factors compared to global scaling factors.

\section*{Data Availability}

The compiled data and meta analysis used in this study is presented in the article and within the supplementary information. 

Specifically, the compiled data is provided as a csv file with each row containing the level of theory, basis set, scaling factor, statistical figure where available, spectral region for fitting, benchmark dataset and source reference. 

We also provide a PDF file detailing the acronyms in the data csv file, as well as some additional analysis highlighting the large errors present in the 99f/26mol data set due to the inclusion of radical species.

\section*{Acknowledgements} 
We thank Anna-Maree Syme for the very insightful discussions on this project. 

% This research was undertaken with the assistance of resources from the National Computational Infrastructure (NCI Australia), an NCRIS enabled capability supported by the Australian Government.

The authors declare no conflicts of interest.

% \bibliography{MoreRefs}
% \bibliographystyle{apsrev}

\end{document}